\documentclass[twocolumn,aps,prb,10pt,amsmath,amssymb,superscriptaddress,nofootinbib,longbibliography]{revtex4-2}

\usepackage{graphicx}
\usepackage{dcolumn}
\usepackage{bm}
\usepackage[dvipsnames]{xcolor}

\usepackage{tabu}
\usepackage{xcolor}

\usepackage{hyperref}
\usepackage{outlines}
\usepackage{multirow}
\usepackage{url}
\usepackage{physics}
\usepackage{pifont}

\begin{document}

\title{Trilinear coupling and toroidicity in multiferroics}

\author{Andrea Urru}
 \email{au168@physics.rutgers.edu}
 \affiliation{Department of Physics \& Astronomy, Rutgers University,
Piscataway, New Jersey 08854, USA}
\affiliation{SISSA -- Scuola Internazionale Superiore di Studi Avanzati, via Bonomea 265, I-34136, Trieste, Italy}
\author{Alessio Filippetti}
\affiliation{Dipartimento di Fisica, Universit\`a di Cagliari, Cittadella Universitaria, I-09042 Monserrato (CA), Italy}
\affiliation{Consiglio Nazionale delle Ricerche, Istituto Officina dei Materiali, CNR-IOM, Cagliari, Cittadella Universitaria, Monserrato 09042-I (CA), Italy.}
\author{Jorge \'I\~niguez-Gonz\`alez}
  \affiliation{Luxembourg Institute of Science and Technology (LIST), Smart Materials Unit, Avenue des Hauts-Fourneaux 5, L-4362 Esch/Alzette, Luxembourg}
  \affiliation{Department of Physics and Materials Science, University of Luxembourg, 41 Rue du Brill, L-4422 Belvaux, Luxembourg}
\author{Vincenzo Fiorentini}
\affiliation{Dipartimento di Fisica, Universit\`a di Cagliari, Cittadella Universitaria, I-09042 Monserrato (CA), Italy}
\affiliation{Institute for Materials Science and Max Bergmann Center for Biomaterials, TU Dresden, D-01062 Dresden, Germany}

\date{\today}

\begin{abstract}
Magnetoelectric responses are related in general to magnetic multipoles, and in particular the off-diagonal linear  response is  proportional to the toroidization order parameter. In multiferroics with three order parameters (polarization, magnetization, and toroidization), such response turns out to be also proportional to the trilinear coupling between the order parameters. Here we explore this scenario, discussing the effects of such trilinear coupling within a Landau theory framework, and using \textit{ab initio} calculations to discuss a specific model system for this effect, namely the recently predicted three-order-parameter multiferroic metal Bi$_5$Mn$_5$O$_{17}$.
\end{abstract}

\maketitle

\section{Introduction}

The response to external perturbations (e.g., electromagnetic fields, strain, etc.) is crucial to identify materials  applications, and provides guidance in the design of new materials with specific functionalities. In the present work we deal with the magnetoelectric (ME) response to electric and magnetic fields.  Magnetoelectricity occurs, among others, in multiferroic materials combining polar and magnetic order, and was first proposed theoretically for Cr$_2$O$_3$ in 1960 by Dzyaloshinskii\cite{dzyaloshinskii-jetp60} and soon after identified experimentally by Astrov\cite{astrov-jetp60}. 

Since the early 2000s, multiferroic materials and the related ME effects have experienced a renaissance\cite{hill-jphyschemb00,hill-jmagmagmat02,spaldin-science05,spaldin-natmater19}, fostered by the discovery of promising materials such as  BiFeO$_3$ \cite{wang-science03} and YMnO$_3$ \cite{vanaken-natmat04}. Increasing effort has been directed  at identifying an order parameter of microscopic origin for linear magnetoelectricity. This has been identified in the so-called \textit{ME multipoles} \cite{spaldin-prb13} appearing in the expansion of the interaction energy between a non-uniform magnetic field $\mathbf{H}(\mathbf{r})$ and the magnetization density $\boldsymbol{\mu}(\mathbf{r})$
\begin{eqnarray}
    E &= & -  \int_{\Omega} \boldsymbol{\mu}(\mathbf{r}) \cdot \mathbf{H}(\mathbf{r}) \, d^3 \mathbf{r} 
    \label{eq:multipoles}\\
    &= & - \left[\int_{\Omega} \boldsymbol{\mu}(\mathbf{r}) \cdot \mathbf{H}(0) \, d^3 \mathbf{r} 
+ \int_{\Omega} r_{i} \mu_{j} (\mathbf{r}) \partial_{i} H_{j}(0) \, d^3 \mathbf{r} + \ldots \right] \nonumber 
\end{eqnarray}
with $\Omega$ the unit cell volume and a sum over the repeated indices $i$ and $j$ implied. In Eq.\eqref{eq:multipoles}, the ME multipoles appear as the first-order moments of the magnetization density $\mathcal{M}_{ij}$=$\int_{\Omega} r_{i} \mu_{j} (\mathbf{r}) \, d^3 \mathbf{r}$. They have a one-to-one link with the linear ME tensor $\alpha_{ij}$: if a component of the multipole tensor is non-zero, the corresponding component of the ME tensor is also non-vanishing. 

Among  these multipoles, here we focus on the \textit{toroidal moment}, a vector quantity corresponding to the anti-symmetric part of $\mathcal{M}_{ij}$ and defined as
\begin{equation}
    \mathbf{t} = \frac{1}{2} \int_{\Omega} \mathbf{r} \times \boldsymbol{\mu} (\mathbf{r}) \, d^3 \mathbf{r}.
    \label{eq:toroidal}
\end{equation}
Given its definition, the toroidal moment appears in the off-diagonal entries of $\mathcal{M}_{ij}$ and it is thus associated with the off-diagonal linear ME response. 

Such correspondence, justified by symmetry, can be further understood  phenomenologically by means of a Landau free energy expansion in the order parameters, polarization $\mathbf{P}$, magnetization $\mathbf{M}$, and toroidization $\mathbf{T}$ (defined as the toroidal moment per unit volume). The simplest free energy  describing a transition from a non-toroidal to a toroidal state and, at the same time, capturing the mutual coupling among the three order parameters\cite{ederer-prb07,sannikov-jetp97} is 
\begin{equation}
\begin{split}
    \mathcal{F} &= a_1 P^2 + a_2 M^2 + a_3 T^2 + \beta \mathbf{T} \cdot (\mathbf{P} \times \mathbf{M}) \\ & \ \ \ - \mathbf{P} \cdot \mathbf{E} - \mathbf{M} \cdot \mathbf{B},
    \end{split}
    \label{eq:landau}
\end{equation}
where $\mathbf{E}$ and $\mathbf{B}$ are the applied electric and magnetic fields, respectively, and $\beta$ is the trilinear coupling {coefficient}. After computing the equilibrium values of $\mathbf{P}$ and $\mathbf{M}$ by minimizing the free energy, it is possible to identify\cite{ederer-prb07} an off-diagonal ME response proportional to the toroidization $\mathbf{T}$, $$ \alpha_{ij} = - \frac{\beta}{a_1 a_2} \sum_k \epsilon_{ijk} T_k,$$ with  $\epsilon_{ijk}$ the antisymmetric Levi-Civita tensor. Remarkably, the ME response is proportional to the trilinear coupling $\beta$ as well. While it is well known that an off-diagonal ME response implies the presence of a net toroidization $\mathbf{T}$ and vice versa, the fact that $\mathbf{T}$ itself is coupled trilinearly to $\mathbf{P}$ and $\mathbf{M}$ in simultaneously ferromagnetic and ferroelectric multiferroics has not been often emphasized.\cite{ederer-prb07,sannikov-jetp97,sannikov-jetp01}

Despite being expected from  Landau free-energy arguments, the presence of a trilinear coupling in realistic materials has not been explored before by direct calculation, mainly because of the scarcity of materials with polarization and magnetization primary order parameters. In this work we fill this gap reporting a trilinear coupling in the recently studied material Bi$_5$Mn$_5$O$_{17}$ (BiMO henceforth).\cite{urru-natcomm2020,belviso-inorgchem2019} BiMO is a ME multiferroic metal, with spontaneous magnetization and polarization in its ground-state. It is predicted to exist in two nearly degenerate ground-states, accessible at room temperature, both displaying a sizable net toroidization as a third order parameter.\cite{urru-natcomm2020} As such, BiMO is ideal to explore the existence and nature of a trilinear coupling term. 
Here, we elaborate on a more general version of the Landau free energy of Eq.\eqref{eq:landau} and discuss how a trilinear coupling affects the energetics, with particular attention to the consequences on the primary (magnetic or electric) orders switching. We then show that BiMO indeed has a trilinear coupling among polarization, magnetization, and toroidization. 

The  paper is organized as follows. In Section \ref{sec:toroidal} we recap some important features of the toroidal moment in solids, which are relevant for its calculation in BiMO. In Section \ref{sec:results} we analyze a Landau-like free energy accounting for trilinear coupling, and we present our results based on \textit{ab initio} calculations demonstrating its presence in BiMO. 

\section{Toroidal moment in extended periodic systems}
\label{sec:toroidal}

Because it contains the position $\mathbf{r}$, which breaks periodic boundary conditions, the toroidal moment is inherently multi-valued, similarly to electrostatic polarization.\cite{king-smith-prb93,resta-ferro92,resta-rmp94,spaldin-jssc2012} We rewrite the integral\cite{spaldin-prb13} in Eq.\eqref{eq:toroidal} as 
\begin{equation}
\begin{split}
    \mathbf{t} &= \frac{1}{2} \sum_{i} \mathbf{r}_{i} \times \mathbf{m}_{i} + \frac{1}{2} \sum_{i} \int_{\Omega_{i}} (\mathbf{r} - \mathbf{r}_i) \times \boldsymbol{\mu} (\mathbf{r}) \, d^3 \mathbf{r},\label{eq:lm-as} \\
    &= \mathbf{t}^{\text{LM}} + \mathbf{t}^{\text{AS}}
    \end{split}
\end{equation}
where $i$ labels the atoms, $\Omega_i$ is the atomic sphere centered on atom $i$, and $\mathbf{m}_i$=$\int_{\Omega_i} \boldsymbol{\mu} (\mathbf{r}) d^3 \mathbf{r}$ is the magnetic dipole moment of atom $i$, with  the implicit and usually acceptable assumption that  magnetic contributions from interstitial regions are negligible.  The first term in Eq.\eqref{eq:lm-as} is usually called the \textit{local moment} (LM) contribution, whereas the second is referred to as the \textit{atomic-site} (AS) contribution. This separation of LM and AS contributions is relevant because the multi-valuedness and possible origin dependence of $\mathbf{t}$ only affect the LM term. The AS contribution is single-valued and origin-independent, which makes it easier to compute, and to link the toroidal moment to material properties by symmetry.\cite{spaldin-prb13} Furthermore, we mention that recently the distinction between the LM and AS terms has become even more fundamentally relevant in the context of surface magnetism.\cite{weber-prx24}

The  magnitude of the toroidal moment is usually dominated by the LM contribution, which will therefore be our focus here. The origin dependence of the LM term for ferromagnetic systems\cite{ederer-prb07} can be removed by rewriting the magnetic moments as 
$$\mathbf{m}_i=(\mathbf{m}_i-\sum_j\mathbf{m}_j/N) +\sum_j\mathbf{m}_j/N =\mathbf{m}_{0,i}+\widetilde{\mathbf{m}},$$ 
the sum of a  compensated part $\mathbf{m}_{0,i}$=$\mathbf{m}_i - \mathbf{m}/N$ (with $\mathbf{m} = \sum_i \mathbf{m}_i$  the net magnetic moment per cell and $N$ the number of magnetic atoms) and an uncompensated, ferromagnetic part
$\widetilde{\mathbf{m}}$=$\mathbf{m}/N.$
The toroidal moment can then be separated into an origin-independent part $\mathbf{t}_0$ and an origin-dependent part $\widetilde{\mathbf{t}}$, given by:
\begin{align}
    \mathbf{t}_0 &= \frac{1}{2} \sum_{i} \mathbf{r}_i \times \mathbf{m}_{0,i}; \ \ \  
    \widetilde{\mathbf{t}} = \frac{1}{2} \overline{\mathbf{r}} \times \mathbf{m},
\end{align}
with $\overline{\mathbf{r}}$=$\sum_i \mathbf{r}_i / N$ the average position of magnetic atoms. Being origin-dependent, $\widetilde{\mathbf{t}}$ is clearly physically irrelevant and in the remainder of this work we remove it by placing the origin of the reference framework at $\overline{\mathbf{r}}$. 

\section{Results}
\label{sec:results}

\subsection{Landau free energy expansion}
\label{sec:trilinear}

Consider a system which presents a phase with three distinct order parameters, namely an electrostatic polarization $\mathbf{P}$, a magnetization $\mathbf{M}$, and a toroidization $\mathbf{T}$. We model it by a Landau-like free-energy expansion in powers of the order parameters up to the fourth order around the origin $(P,M,T)$=$(0,0,0)$. 
To simplify  further, we neglect spatial anisotropy, and the 3D nature of the order parameters, only considering their magnitudes as variables in our theory. Introducing a compact notation identifying the order parameters $P$, $M$, and $T$ by the array $\eta_2$=$(P, M, T)$, the free energy reads 
\begin{equation}
    \label{eq1}
    \mathcal{F}(P,M,T) = \mathcal{F}_0 + \eta_2 A \eta_2^\intercal + \eta_4 C \eta_4^\intercal + \beta M P T,
\end{equation}
where $\mathcal{F}_0$ is the free energy at $(P,M,T)$=$(0,0,0)$, $\eta_4$=$(P^2, M^2, T^2)$ contains the squared order parameters, $A$ is a  diagonal 3$\times$3 matrix, $C$ is a symmetric 3$\times$3 matrix,
and $\beta$ is the coefficient for the trilinear coupling of $P$, $M$, and $T$.
In Eq.\eqref{eq1},  third-order terms, second-order mixed terms, and mixed fourth-order terms of type $TP^3$ (with any permutation of order parameters) are forbidden because they break either space or time inversion. In this approximation, $C$ must be positive definite for $\mathcal{F}$ to have a lower bound. Consequently $\mathcal{F}$ shows non-trivial minima if any of the diagonal elements $a_1$, $a_2$, $a_3$  of $A$ 
is negative. 

We assume that above some critical temperature $T_c$ the system described by Eq.\eqref{eq1} is in a high-symmetry state with no order parameter, i.e. the free energy landscape has only one minimum at $(P, M, T)$=$(0, 0, 0)$. On the other hand, we assume that below $T_c$ at least one among  $a_1$, $a_2$, $a_3$ is negative, so that the free energy landscape has non trivial minima and the system can occupy a low-symmetry state with non-vanishing order parameters $P$, $M$, and $T$. 

Next, we discuss the allowed stationary points of the free energy landscape in the two cases $\beta$=0 and $\beta$$\ne$0, i.e., in the absence or presence of a trilinear coupling, respectively.
\subsubsection{\texorpdfstring{The $\beta$=0}{beta} case}
The critical points of $\mathcal{F}(P,M,T)$, Eq.\eqref{eq1}, correspond to the condition $\grad \mathcal{F}$=$0$, i.e.: 
\begin{subequations}
\begin{align}
    P \left( 2 a_1 + 4 c_{11} P^2 + 4 c_{12} M^2 + 4 c_{13} T^2 \right) = 0, \\
    M \left( 2 a_2 + 4 c_{12} P^2 + 4 c_{22} M^2 + 4 c_{23} T^2 \right) = 0, \\ 
    T \left( 2 a_3 + 4 c_{13} P^2 + 4 c_{23} M^2 + 4 c_{33} T^2 \right) = 0.
\end{align}
\end{subequations}
Each equation has three solutions, thus there are overall 27 stationary points, listed for completeness in Table \ref{t1}. Note that by construction the model describes as a stationary state every ordered phase with one, two, or three distinct order parameters, whose values we identify by $\widetilde{P}$, $\widetilde{M}$, and $\widetilde{T}$, respectively. Among the stationary points there is also the trivial solution $(\widetilde{P}, \widetilde{M}, \widetilde{T})$=$(0, 0, 0)$, which shall not be a minimum in the low-temperature phase. 

\begin{table}[t]
\caption{List of stationary points of $\mathcal{F}$, their behavior under inversion and time-reversal symmetries, and their multiplicity, for the $\beta$=0 case.}
    \centering
    {\renewcommand{\arraystretch}{1.2}
    \begin{tabular}{|c|c|c|c|}
        \hline
        Stationary point & Inversion & Time-reversal & Multiplicity \\
        \hline
        (0, 0, 0) & Y & Y & 1 \\
        \hline
        ($\pm \widetilde{P}$, 0, 0) & N & Y & 2 \\
        \hline
        (0, $\pm \widetilde{M}$, 0) & Y & N & 2 \\
        \hline
        (0, 0, $\pm \widetilde{T}$) & N & N & 2 \\
        \hline
        ($\pm \widetilde{P}$, $\pm \widetilde{M}$, 0) & N & N & 4 \\
        \hline
        ($\pm \widetilde{P}$, 0, $\pm \widetilde{T}$) & N & N & 4 \\
        \hline
        (0, $\pm \widetilde{M}$, $\pm \widetilde{T}$) & N & N & 4 \\
        \hline
        ($\pm \widetilde{P}$, $\pm \widetilde{M}$, $\pm \widetilde{T}$) & N & N & 8 \\
        \hline
        \end{tabular}     
         }
    \label{t1}
\end{table}

To ascertain the nature of the stationary points, a second-derivative test based on the Hessian matrix of $\mathcal{F}$, $\mathcal{H}_{ij}$=$\partial_i \partial_j \mathcal{F}$, with $i, j$ identifying the variables $P$, $M$, and $T$, is required. The Hessian matrix  of $\mathcal{F}$ (Eq.\eqref{eq1}) reads
\begin{equation}
    \mathcal{H} = \begin{pmatrix} h_{11} & 8 c_{12} P M & 8 c_{13} P T \\ 8 c_{12} P M & h_{22} & 8 c_{23} M T \\ 8 c_{13} P T & 8 c_{23} M T & h_{33} \end{pmatrix},
\end{equation}
where 
\begin{subequations}
\begin{align}
    h_{11} &= 2 a_1 + 12 c_{11} P^2 + 4 c_{12} M^2 + 4 c_{13} T^2, \\ 
    h_{22} &= 2 a_2 + 12 c_{22} M^2 + 4 c_{12} P^2 + 4 c_{23} T^2, \\ 
    h_{33} &= 2 a_3 + 12 c_{33} T^2 + 4 c_{13} P^2 + 4 c_{23} M^2.
\end{align}
\end{subequations}
For the trivial solution, the Hessian is diagonal, 
\begin{equation}
    \mathcal{H} = 2 \begin{pmatrix} a_1 & 0 & 0 \\ 0 & a_2 & 0 \\ 0 & 0 & a_3 \end{pmatrix},
\end{equation}
implying that below $T_{\text{c}}$ the point $(P, M, T)$=$(0, 0, 0)$ is not a minimum if any of the coefficients $a_1$, $a_2$, or $a_3$ is negative. The nature of the general stationary points in Table \ref{t1} will be not equally straightforward to analyze. Because $\mathcal{F}$ is required to be bounded from below, however, we expect at least a global minimum, and given the large number of parameters in the model, it is likely that the stationary points will include local minima, which are determined by the three eigenvalues of the Hessian being all positive.
Thus in the following we assume we are dealing with minima (local or global).

We now consider  the minima characterized by three non-vanishing order parameters. It is worth noting that in the absence of a trilinear coupling, the free energy $\mathcal{F}$ and its hessian matrix are symmetric under sign reversal of any order parameter ($P \rightarrow -P$, etc.). Therefore, each of the eight stationary points in the family $(\pm \widetilde{P}, \pm \widetilde{M}, \pm \widetilde{T})$ has the same free energy and the same behavior, i.e., if one point is a local or global minimum, so is each of the remaining seven. In particular, this implies that the toroidization $T$ is independent of $P$ and $M$, i.e., it is not uniquely determined by $P$ and $M$, since both states $(P, M, T)$ and $(P, M, -T)$ are minima with the same free energy. 

\subsubsection{\texorpdfstring{The $\beta$$\ne$0}{beta2} case}

For $\beta$$\ne$0 the condition $\grad \mathcal{F}$=0, i.e. 
\begin{subequations}
\begin{align} 
    P \left( 2 a_1 + 4 c_{11} P^2 + 4 c_{12} M^2 + 4 c_{13} T^2 \right) + \beta M T &= 0, \label{eq:stationary} \\
    M \left( 2 a_2 + 4 c_{12} P^2 + 4 c_{22} M^2 + 4 c_{23} T^2 \right) + \beta P T &= 0, \\ 
    T \left( 2 a_3 + 4 c_{13} P^2 + 4 c_{23} M^2 + 4 c_{33} T^2 \right) + \beta P M &= 0, \label{eq:t_stationary} 
\end{align}
\end{subequations}
provides the stationary points summarized in Table \ref{t2}.
\begin{table}[ht]
\caption{Stationary points of $\mathcal{F}$ for $\beta$$\ne$0 and their behavior under inversion and time-reversal, and multiplicity. In the bottom row, we give one representative for each family of four states, called $\mathcal{S}_1$ and $\mathcal{S}_2$ in the main text - the remaining three stationary points of the family are obtained from the representative one by reversing any two order parameters.}
    \centering
    {\renewcommand{\arraystretch}{1.2}
    \begin{tabular}{|c|c|c|c|}
        \hline
        Stationary point & Inversion & Time-reversal & Multiplicity \\
        \hline
        (0, 0, 0) & Y & Y & 1 \\
        \hline
        ($\pm \widetilde{P}$, 0, 0) & N & Y & 2 \\
        \hline
        (0, $\pm \widetilde{M}$, 0) & Y & N & 2 \\
        \hline
        (0, 0, $\pm \widetilde{T}$) & N & N & 2 \\
        \hline
        ($\widetilde{P}$, $\widetilde{M}$, $\widetilde{T}$) [$\mathcal{S}_1$] & \multirow{3}{*}{N} & \multirow{3}{*}{N} & \multirow{3}{*}{4} \\
        or & & & \\
        ($- \widetilde{P}$, $- \widetilde{M}$, $- \widetilde{T}$) [$\mathcal{S}_2$] & & & \\        \hline
        \end{tabular}     
         }
    \label{t2}
\end{table}
Similarly to the $\beta$=0 case, these include the trivial solution $(\widetilde{P}, \widetilde{M}, \widetilde{T})$=$(0, 0, 0)$ and the single-order-parameter solutions $(\pm\widetilde{P}, 0, 0)$, $(0, \pm\widetilde{M}, 0)$, $(0, 0, \pm\widetilde{T})$. For these, the trilinear coupling term vanishes, thus the specific value of $\widetilde{P}$, $\widetilde{M}$, and $\widetilde{T}$ cannot depend on $\beta$, as  expected from the definition of a trilinear coupling describing the interplay between three distinct order parameters. 

At variance with the $\beta$=$0$ case, solutions with two order parameters are not allowed: for instance, if $T$=$0$, from Eq. \eqref{eq:t_stationary} we get $\beta P M$=$0$,  implying that at least one other order parameter vanishes. Also, the number of solutions with three non-vanishing order parameters drops to four, from the eight for the $\beta$=$0$ case. This is because the trilinear coupling $\beta P M T$ is invariant only under an even number of sign reversals. 

To clarify this point further, let us take a stationary solution $(\widetilde{P}, \widetilde{M}, \widetilde{T})$. If we reverse the polarization $\widetilde{P}$, according to the stationarity condition in Eq. \eqref{eq:stationary}, this is still a solution only if either $\widetilde{M}$ or $\widetilde{T}$ reverses as well. As a consequence, only one of the  families 
$$
\mathcal{S}_1\!\!\equiv\!\![(\widetilde{P},\widetilde{M},\widetilde{T}), (\widetilde{P},-\widetilde{M},-\widetilde{T}), (-\widetilde{P},-\widetilde{M},\widetilde{T}), (-\widetilde{P},\widetilde{M},-\widetilde{T})]
$$ and 
$$
 \mathcal{S}_2\!\!\equiv\!\![(-\widetilde{P}, \widetilde{M}, \widetilde{T}), (\widetilde{P}, -\widetilde{M}, \widetilde{T}), (\widetilde{P}, \widetilde{M}, -\widetilde{T}), (-\widetilde{P}, -\widetilde{M}, -\widetilde{T})]
 $$
 is stationary. Below, we simplify the notation identifying each state only by the sign of the order parameters: for instance, $(+, -, +)$ is $(\widetilde{P}, -\widetilde{M}, \widetilde{T})$.

The discussion so far indicates that one can ascertain the presence or absence of a trilinear order-parameter coupling  by picking one member of each family of four-fold degenerate states, e.g., $(+, +, +)$ and $(+, +, -)$, and verify with a direct calculation whether they are degenerate in energy or not. In the following we show with a specific case study that all the states in the  $\mathcal{S}_2$ family are stable, energy-degenerate, and with the same order parameters, modulo signs, whereas  states in the $\mathcal{S}_1$ family are not stationary points, confirming the existence of a trilinear coupling.

\subsection{\texorpdfstring{B\lowercase{i}$_5$M\lowercase{n}$_5$O$_{17}$}{BiMO} as a case study}

\begin{figure}[t]
\centering\includegraphics[width=\columnwidth]{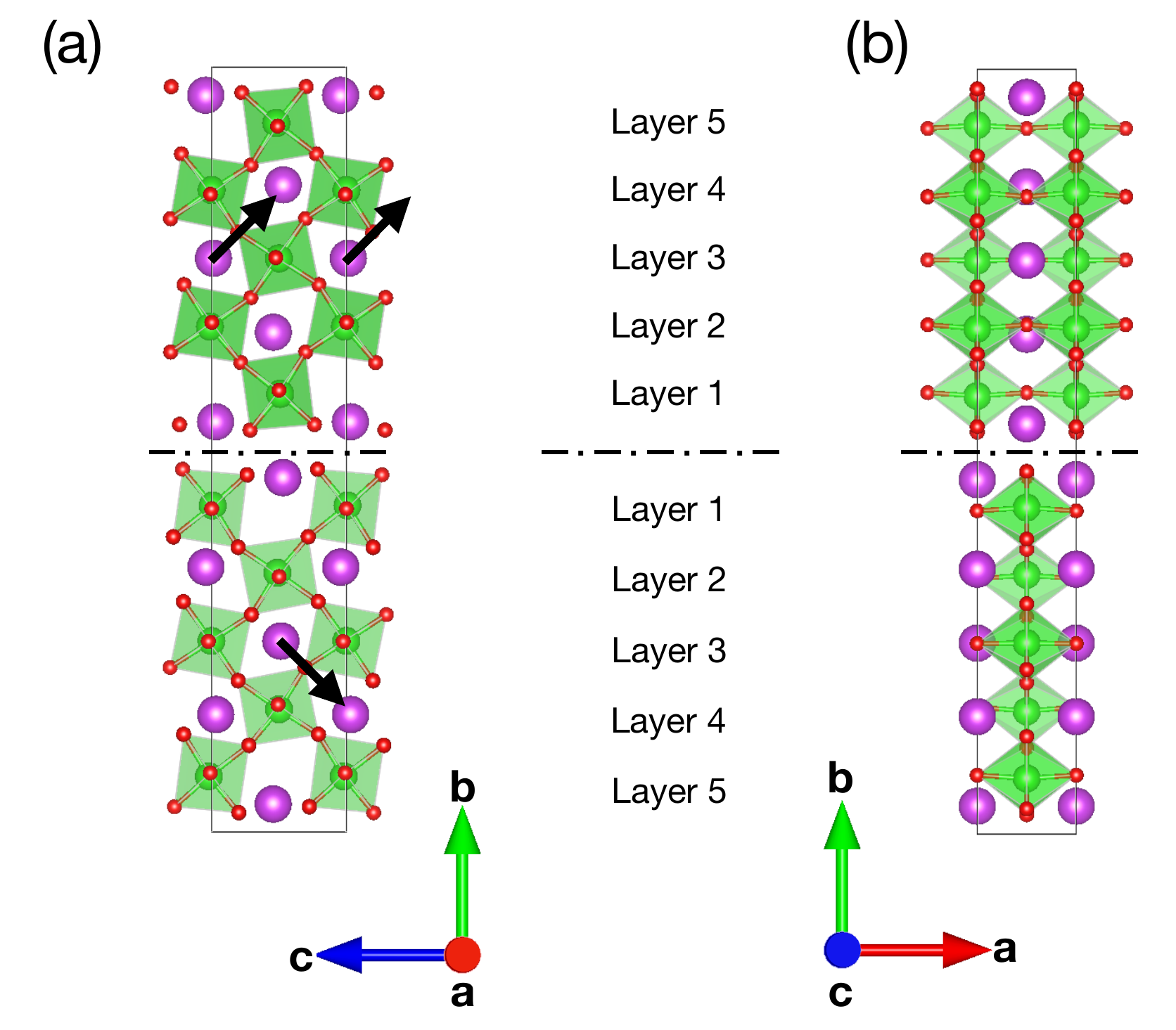}
\caption{Crystal structure of BiMO for ground state C, with polarization along $\mathbf{c}$. (a) Side view perpendicular to $\textbf{a}$. Black arrows identify the main atomic displacements of the $B_{3u}$ mode of the reference P$mnn$ phase, responsible for the polarization. (b) Side view perpendicular to $\textbf{c}$. The structure is made up of two blocks of five layers each, separated -- and related to each other -- by a $n$ glide plane, identified by a dashed-dotted line, with fractional translation $(1/2, 0, 1/2)$. Layers related by the glide plane are identified with the same number.} 
\label{fig:struct}
\end{figure}

BiMO is a recently predicted\cite{urru-natcomm2020} multiferroic metal belonging  to the family of Carpy-Galy layered perovskites A$_n$B$_n$O$_{3n+2}$, with $n$=5. These are compounds made up of blocks of $n$ layers containing corner-sharing distorted BO$_6$ octahedra, stacked along the [110] pseudo-cubic direction, with each block separated from the neighboring one by an A$_2$O$_2$ layer. BiMO is predicted to exist in two different nearly-degenerate ground states with respective magnetic space groups P$m'n2_1'$ and P$m2_1'n'$. As in Ref.\cite{urru-natcomm2020}, we call these states C and B, with reference to their respective polar directions (the {\bf c} and {\bf b} crystal axes).  Fig.\ref{fig:struct} shows the side views of the crystal structure of state C. The layers of the two perovskite-like blocks are symmetry-related in pairs by the glide plane $n$ perpendicular to $\mathbf{b}$ (dashed-dotted line in Fig.\ref{fig:struct}).

As predicted with \textit{ab initio} calculations (see Appendix \ref{sec:details} for the computational details), the two ground states C and B are multiferroic: they both possess simultaneously a polarization $\mathbf{P}$, a magnetization $\mathbf{M}$, and a toroidization $\mathbf{T}$, but with different orientations. In state C we find $\mathbf{P}$$\|${\bf c}, $\mathbf{M}$$\|${\bf b}, and $\mathbf{T}$$\|${\bf a}, whereas in state B the order parameters  are rotated:  $\mathbf{P}$$\|${\bf b}, $\mathbf{M}$$\|${\bf a}, and $\mathbf{T}$$\|${\bf c}.
Both ground states are dynamically stable and are obtained by condensing the zone-center unstable phonon modes of a parent centrosymmetric (hence non-polar and non-toroidal) phase with P$mnn$ symmetry. The latter is the reference undistorted structure with respect to which the polarization and toroidization of the acentric multiferroic ground states are computed. Below we focus on state C, but state B behaves similarly. 

\subsubsection{Toroidal moment analysis}

The parent reference phase (R henceforth) P$mnn$ and the low-symmetry state C (P$mn2_1$) are connected by a distortion pattern which decomposes into zone-center phonon modes of R as 
\begin{equation}
    \boldsymbol{\Delta} \mathbf{r}_i = a \mathbf{u}_i^{A_g} + b \mathbf{u}_i^{B_{3u}} 
\end{equation}
using the \texttt{AMPLIMODES} code.\cite{orobengoa-jac09,perez-mato-acsa10} Here $\boldsymbol{\Delta} \mathbf{r}_i$ describes the displacement of atom $i$ as the combination of  two normalized displacement modes belonging to the $B_{3u}$ and $A_g$ irreducible representations of the $mmm$ point group. The $B_{3u}$ mode is responsible for the symmetry lowering of the $\text{R}$$\rightarrow$$\text{C}$ phase transition. Its main component is given by the off-centering of the Bi atoms in the third layer, as shown in Fig. \ref{fig:struct}. On the other hand, the $A_g$ is total-symmetric and we will not consider it further because it does not lower the symmetry of P. From the mode decomposition analysis giving $b$=0.27 and $a$=0.09, $B_{3u}$ emerges as the primary distortion, as expected. 

Next, we compute the toroidal moment by following a distortion path from R to C. The path is constructed from the collective distortion identified by $\boldsymbol{\Delta} \mathbf{r}_i$ and it is scaled by a factor ranging between $-1$ and $+1$. The end values correspond to state C with opposite polarization, whereas $0$  corresponds to the R phase. It is worth noting that, besides a clear difference in crystal structure, the R and C states  also  differ in their magnetic moments. Both  have ferromagnetic Mn magnetic moments along the $\textbf{b}$ direction, but the Wyckoff site symmetry of Mn atoms allows for additional antiferroic canting of the magnetic moments in the plane perpendicular to $\mathbf{b}$, and these cantings differ in the R and C states.  We call $\boldsymbol{\Delta}$$\mathbf{m}_i$ the difference between the magnetic moments of the two phases, and  linearly interpolate it between R and C along the distortion path, similarly to the displacements identified by $\boldsymbol{\Delta} \mathbf{r}_i$. 

We note that the antiferroic cantings are induced by spin-orbit coupling. To describe them, non-collinear calculations with relativistic effects included are needed, as mentioned in Appendix \ref{sec:details}.

\begin{figure}[t]
\centering\includegraphics[width=0.85\columnwidth]{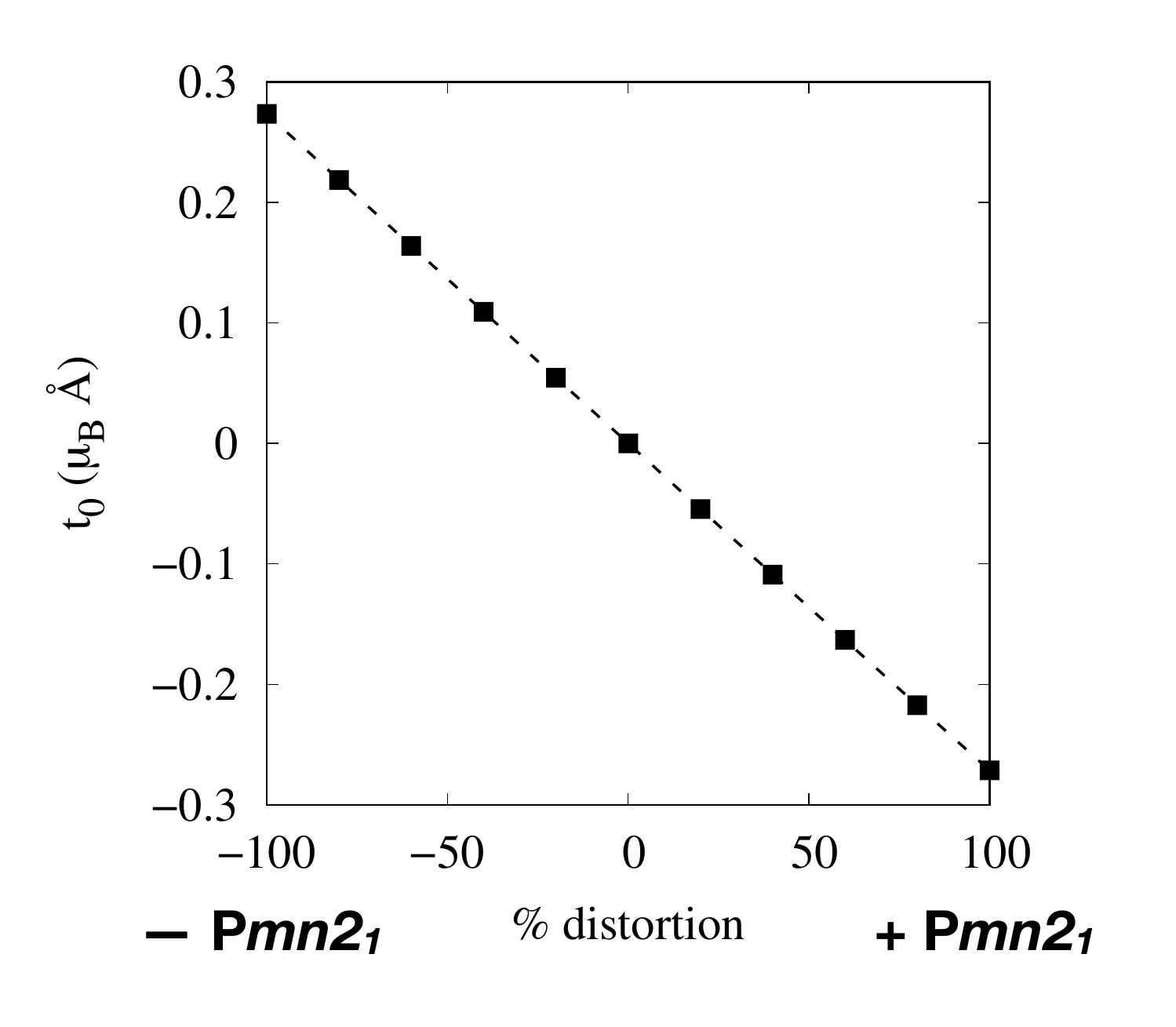}
\caption{Toroidal moment of BiMO along the distortion path from the P$mnn$ to the P$mn2_1$ phase.} 
\label{fig:toroidal}
\end{figure}

In Fig. \ref{fig:toroidal} we show the toroidal moment of BiMO along the distortion path connecting low-symmetry C states with opposite polarization through the high-symmetry phase R. The data show no discontinuity to different branches connected by a toroidal moment increment. Indeed, because of the multi-valuedness of the toroidal moment (Section \ref{sec:toroidal}), the non-toroidal R phase may end up having a net toroidal moment corresponding to a toroidal moment increment. We checked for this in P$mnn$ BiMO, and explicitly removed any toroidal moment increment for all the structures along the distortion path. 

\begin{figure}[t]
\centering\includegraphics[width=\columnwidth]{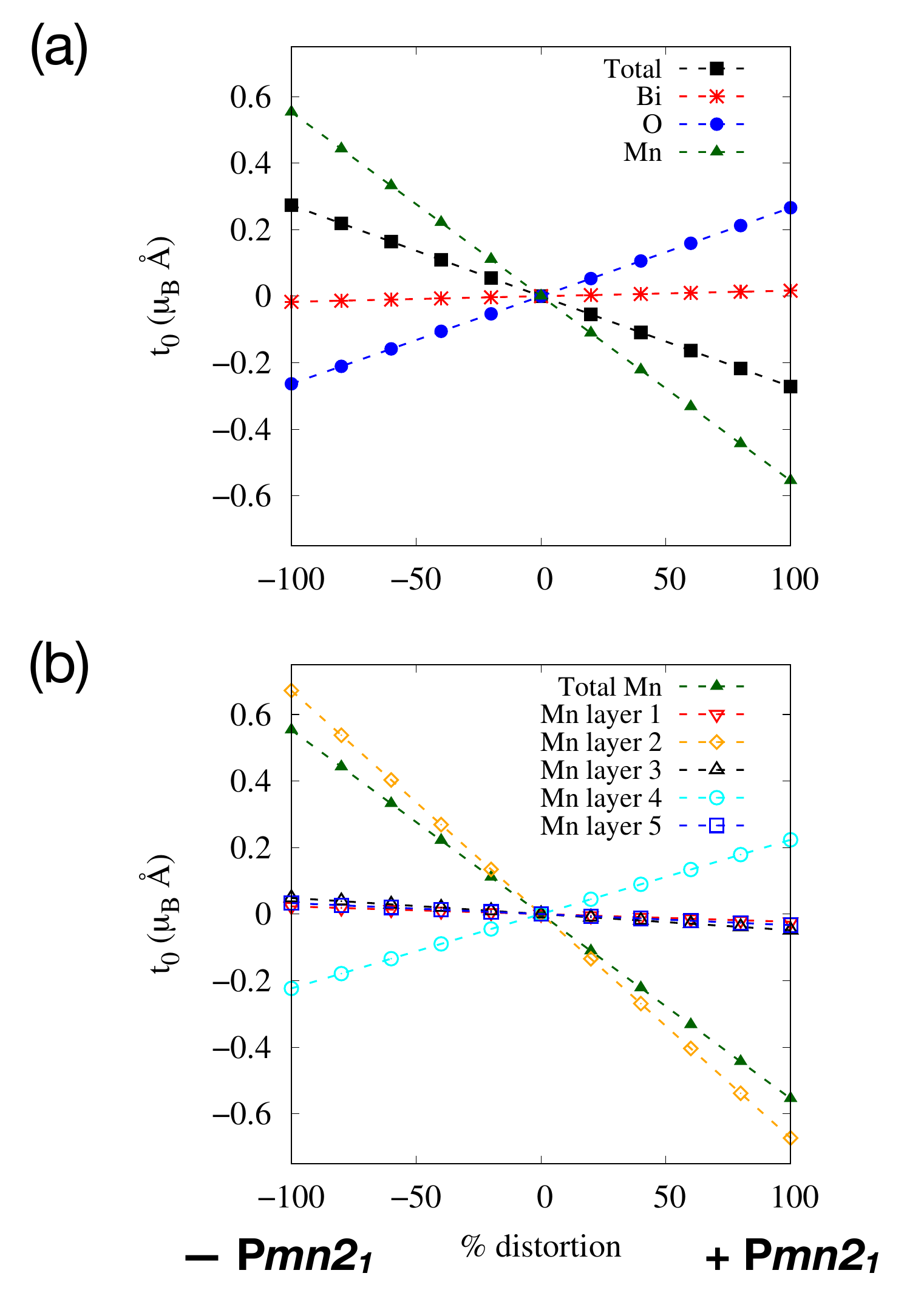}
\caption{Toroidal moment of BiMO: contributions from (a) different atomic species and (b) Mn atoms of each perovskite layer. Layer numbering as in Fig. \ref{fig:struct}}
\label{fig:toroidal_atoms}
\end{figure}

To pinpoint the source of toroidal moment, we decompose it in contributions from each atom separately. The resulting decomposition, Fig.\ref{fig:toroidal_atoms}a, shows that, as expected, Mn atoms are the primary source of toroidal moment.  O atoms contribute a significant compensating toroidal moment (about half that of the Mn's in magnitude and opposite in sign), due to their larger number and despite their induced magnetic moment being small. This reduces the magnitude of the overall toroidal moment  for the C state, which ends up being  about 0.28 $\mu_{\text{B}}$\AA, in the same ballpark as other reported ferrotoroidal compounds (e.g., LiMPO$_4$ with M=Fe, Ni, Mn, Co) \cite{ederer-prb07}. 

In Fig.\ref{fig:toroidal_atoms}(b) we break down the Mn toroidal moments into contributions from the different layers in  each block of the layered structure. The Mn atoms of layer 2, due to their bigger displacement compared to the other Mn's, give the largest contribution to the total toroidal moment. This is partially reduced by the Mn of the fourth layer, with a toroidal moment about 30\% in magnitude compared to the second layer and opposite in sign. The odd layers carry a much smaller toroidal moment.

\subsubsection{Trilinear coupling}

To assess the presence of a trilinear coupling, according to the discussion in Section \ref{sec:trilinear} we have to ascertain whether the two families $\mathcal{S}_1$ and $\mathcal{S}_2$
are both ground states for BiMO or not. This requires checking whether (i) both configurations are stationary and stable and if so (ii) whether they have the same energy or not. If both configurations are ground states then a trilinear coupling $\beta$ is absent, otherwise $\beta\ne0$.

We preliminarily discuss how  one family of configurations is connected to the other. The state C of BiMO discussed earlier has positive polarization and magnetization and negative toroidization. In our notation, this is identified by $(+, +, -)$ and belongs to family $\mathcal{S}_2$ according to our convention (Section \ref{sec:trilinear}). To  reach the configuration $(+, +, +)$, the  toroidicity needs to be reversed. We start by writing the toroidal moment $\mathbf{t}$ of state C in terms of the distortions $\boldsymbol{\Delta} \mathbf{r}_i$ and changes in the magnetic moment, $\boldsymbol{\Delta} \mathbf{m}_i$, relative to the high-symmetry phase. The relevant part of $\mathbf{t}$ is the origin-independent one (as discussed in Section \ref{sec:toroidal}), which reads 
\begin{eqnarray}
    \mathbf{t}_{0} & =& (\mathbf{r}^{(\text{R})}_i + \boldsymbol{\Delta} \mathbf{r}_i) \times (\mathbf{m}^{(\text{R})}_{0, i} + \boldsymbol{\Delta} \mathbf{m}_{0, i})     \label{eq:decomp}\\
    & = & \mathbf{r}^{(\text{R})}_i\times\boldsymbol{\Delta}\mathbf{m}_{0, i} + \boldsymbol{\Delta}\mathbf{r}_i\times\mathbf{m}^{(\text{R})}_{0, i}+\boldsymbol{\Delta}\mathbf{r}_i\times \boldsymbol{\Delta}\mathbf{m}_{0, i},\nonumber
\end{eqnarray}
where a sum over the atoms $i$ is implied,  superscripts $(\text{R})$ refer to the parent reference phase R. Since the R phase is non-toroidal, the  $\mathbf{r}^{(\text{R})}_i$$\times$$\mathbf{m}^{(\text{R})}_{0, i}$ term vanishes and was dropped.

\begin{figure}[t]
\centering\includegraphics[width=\columnwidth]{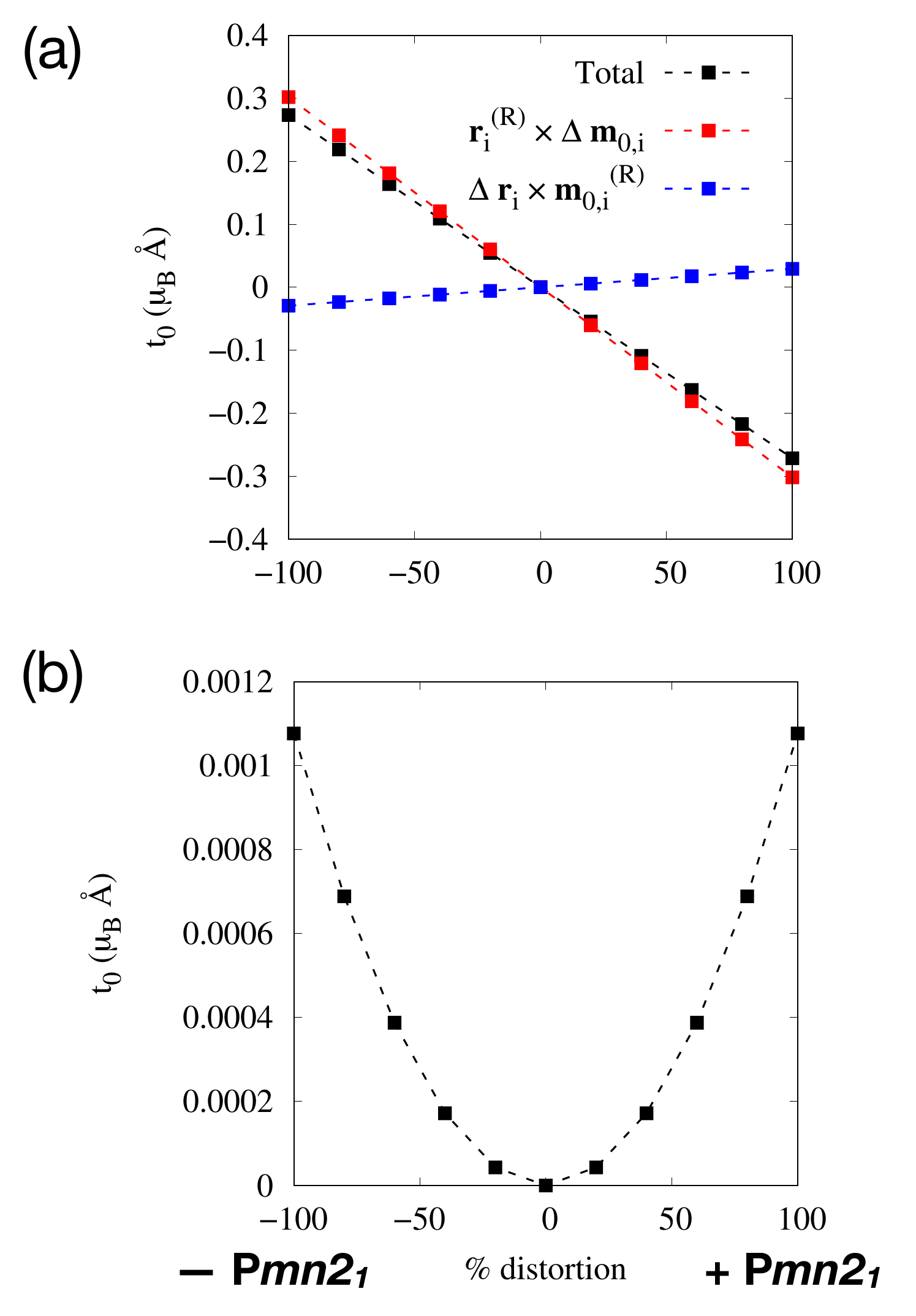}
\caption{Toroidal moment of BiMO: contributions from the terms (a) linear and (b) quadratic in the changes $\boldsymbol{\Delta} \mathbf{r}$ and $\boldsymbol{\Delta} \mathbf{m}$ in positions and magnetic moments, respectively, in Eq.\eqref{eq:decomp}.}
\label{fig:toroidal_decomp}
\end{figure}

As per Eq.\eqref{eq:decomp},  $\mathbf{t}_0$ can be switched simply reversing $\boldsymbol{\Delta} \mathbf{m}_0$ and $\mathbf{m}_0$, without reversing the polarization or the magnetization.\cite{noteM}
In BiMO, this is even more straightforward because $\mathbf{t}_0$ is dominated (see Fig.\ref{fig:toroidal_decomp}a) by the term $\mathbf{r}^{(\text{R})}_i$$\times$$\boldsymbol{\Delta}\mathbf{m}_{0, i}$ which is about 90\% of $\mathbf{t}_0$, while $\boldsymbol{\Delta} \mathbf{r}_i$$\times$$\mathbf{m}^{(\text{R})}_{0, i}$ accounts for about 10\%, and $\boldsymbol{\Delta}\mathbf{r}_i$$\times$$\boldsymbol{\Delta}\mathbf{m}_{0, i}$ is negligible. Thus $\mathbf{t}_0$ is most easily reversed by changing sign to $\boldsymbol{\Delta}\mathbf{m}_{0, i}$. 

We then prepare our system in the $(+, +, +)$ state by reversing $\mathbf{t}_0$ in the way just described, and assess whether this is a ground state configuration. Our calculations show that during self-consistentency the antiferromagnetic components $\boldsymbol{\Delta} \mathbf{m}_{0,i}$ of the magnetic moments tend to switch their sign, and consequently $\mathbf{t}_0$ changes sign with respect to the initial configuration, i.e., tends to go back to the $(+,+,-)$ configuration. This suggests that the $(+, +, +)$ state is not close to a local minimum of the energy landscape and that the $(+, +, -)$ state is stable and energetically favorable. 

To further corroborate our finding, we prepare BiMO in the $(-, -, +)$ state of the $\mathcal{S}_1$ family by reversing all $\boldsymbol{\Delta} \mathbf{r}_i$ and all magnetic moments, including $\boldsymbol{\Delta} \mathbf{m}_{0,i}$. Similarly to the case of $(+, +, +)$,  the $(-, -, +)$ state is not self-consistently stable, and the $\boldsymbol{\Delta} \mathbf{m}_{0,i}$ tend to switch sign leading to the $(-, -, -)$ configuration of the $\mathcal{S}_2$ family. 

We have shown that the $\mathcal{S}_1$ family of states is not a stable configuration and, hence, is not a ground state of BiMO. According to the discussion in Section \ref{sec:trilinear}, this indicates that BiMO has a trilinear coupling among polarization, magnetization, and toroidization. It further suggests that $\mathbf{T}$ is an improper (or secondary) order parameter, as we discuss later, that would not appear in the absence of $\mathbf{P}$ or $\mathbf{M}$, and that would switch only following the switching of either $\mathbf{P}$ or $\mathbf{M}$.

As an additional check, we take other states of the $\mathcal{S}_2$ family, specifically $(+, -, +)$ and $(-, +, +)$, and check that they are ground states. We summarize our results in Table \ref{tab:s2_family} and describe them in detail next. We start from the $(+, -, +)$ state, which  is reached from $(+, +, -)$ by reversing all the magnetic moments, including their antiferromagnetic components. The resulting configuration is stable, has the same polarization as the $(+, +, -)$ state (5.03 $\mu$C/cm$^2$), opposite magnetization (by construction) and opposite toroidal moment: $- 0.28 \mu_{\text{B}}$\AA,  for $(+, +, -)$ and 
0.3 $\mu_{\text{B}}$\AA\, for $(+, -, +)$. The two states differ in energy by about 1 $\mu$eV/Mn atom, the typical resolution of non-collinear magnetic calculations. Next, we address the $(-, +, +)$ state which, unlike the previous one, is obtained from $(+, +, -)$ by reversing the distortions $\boldsymbol{\Delta} \mathbf{r}_i$ and the changes in magnetic moment $\boldsymbol{\Delta} \mathbf{m}_{0,i}$. We emphasize again that $\boldsymbol{\Delta} \mathbf{m}_{0,i}$ are relative to the antiferromagnetic components of the magnetic moments, thus their reversal does not switch the total magnetization. We check that this configuration has indeed a polarization of 
$- 5.03  \mu$C/cm$^2$, opposite to that of $(+, +, -)$. Also the toroidal moment (0.29 $\mu_{\text{B}}$\AA) switches sign compared to $(+, +, -)$, confirming that this is indeed the $(-, +, +)$ configuration. Its energy is the same as for $(+, +, -)$ within 2 $\mu$eV/Mn atom, again within computational resolution. We do not check the fourth state of the $\mathcal{S}_2$ family, namely $(-, -, -)$, because it is related to $(-, +, +)$ in the same way as $(+, -, +)$ is related to $(+, +, -)$, thus we expect it to behave similarly. 

\begin{table}[t]
\caption{Polarization, magnetization, toroidal moment, and energy difference with respect to the $(+, +, -)$ state for three states of the $\mathcal{S}_2$ family. The reported values are obtained with DFT calculations.}
    \centering
    {\renewcommand{\arraystretch}{1.2}
    \begin{tabular}{|c|c|c|c|c|}
    \hline
    State & $P$($\mu$C/cm$^2$) & $M$($\mu_{\text{B}}$/Mn) & $t_0$($\mu_{\text{B}}$\AA) & $\Delta E$($\mu$eV/Mn) \\
        \hline
        $(+, +, -)$ & $\phantom{+}$5.03 & $\phantom{+}$3.11 & $-$0.28 & 0 \\
        \hline
        $(+, -, +)$ & $\phantom{+}$5.03 & $-$3.11 & $\phantom{+}$0.30 & 1 \\
        \hline
        $(-, +, +)$ & $-$5.03 & $\phantom{+}$3.11 & $\phantom{+}$0.29 & 2 \\
        \hline
        \end{tabular}     
         }
    \label{tab:s2_family}
\end{table}

Our analysis thus shows that states of the $\mathcal{S}_2$ family are stable, energy-degenerate, and with the same order parameters (modulo  signs), whereas those of the $\mathcal{S}_1$ family are not stable. This confirms that in BiMO the polarization, magnetization, and toroidization are related by a trilinear coupling.

Overall, in BiMO the toroidization appears because of the space-inversion and time-reversal symmetry breaking, driven by $P$ and $M$, respectively. As such, $T$ is an improper order parameter, that would not occur in the absence of $P$ or $M$. This is consistent with the presence of a trilinear coupling, which dictates that the order parameters switch in pairs -- in other words, $T$ follows $P$ and $M$ and switches only if $P$ or $M$ switch. Being a secondary order parameter, it is reasonable to assume that $T$ is small and that it is coupled to $P$ and $M$ only through the trilinear term. The resulting simplified Landau model, according to Eq. \eqref{eq:t_stationary}, gives $T = - \frac{\beta}{2 a_3} P M$, i.e., the toroidization is uniquely determined by the primary orders $P$ and $M$. This is a simplified restatement of the relation $\mathbf{T} \propto \mathbf{P} \times \mathbf{M}$ that has been pointed out in Refs.\ \cite{sawada-prl05,yamasaki-prl06}. We note that from the relationship among the three order parameters, it is possible to compute $\beta/a_3$. For BiMO, we find that $\beta/a_3 \approx 3 \times 10^{-5}$ cm$^3$/C.

\section{Summary and conclusions}
\label{sec:conclusions}

In summary, motivated by the relation of the linear magnetoelectric response with the toroidization and the trilinear coupling, we have shown how toroidization, polarization, and magnetization can be trilinearly coupled. We have discussed a Landau theory picture of the phenomenon, and showed that this scenario is realized in the three-order-parameter multiferroic metal Bi$_5$Mn$_5$O$_{17}$.

Our findings have interesting consequences on multi-order-parameter switching. As explained, if a trilinear coupling is present, only a single family of four states is stationary, with the states being related to each other by reversing an even number of order parameters. This suggests that, by applying the appropriate conjugate field, two order parameters will switch at the same time. For example, when an electric field is applied, either magnetization or toroidization will switch together with polarization. 

In a realistic setting, which order parameter reverses together with the one conjugate to the applied field will depend on the details of the switching path and on the energy barrier between the states involved (which, we note, will also include those of the B ground state degenerate with the C state discussed here). Overall, materials with three order parameters coupled by a trilinear term are an interesting platform to explore, because they generalize the concept of traditional multiferroics in  ways that call for additional theoretical and, especially, experimental investigation.

\acknowledgments
A.U. acknowledges support from the Abrahams Postdoctoral Fellowship of the Center for Materials Theory at Rutgers University. A.F. acknowledges project PRIN 2022 "TOTEM", grant No. F53D23001080006, funded by Italian Ministry of University and Research (MUR); project PNRR-PRIN 2022 "MAGIC", grant No. F53D23008340001, funded by EU; project  Network 4 Energy Sustainable Transition "NEST",   award number PE0000021, funded under the National Recovery and Resilience Plan (NRRP), Mission 4, Component 2, Investment 1.3 - Call for tender No. 1561 of 11.10.2022 of MUR, funded by NextGenerationEU. J. Í.-G. acknowledges the financial support from the Luxembourg National Research Fund (FNR) through grant C21/MS/15799044/FERRODYNAMICS. V.F. acknowledges the University of Cagliari for a sabbatical leave and the Chair for materials science and nanotechnology of TU Dresden for welcoming him as a senior Dresden fellow.
Computational facilities have been provided by SISSA through its Linux Cluster and ITCS.

\appendix
\section{Computational details}
\label{sec:details}

\textit{Ab initio} DFT calculations are performed within the local density approximation (LDA), as implemented in \texttt{Quantum ESPRESSO},\cite{giannozzi-jpcm09,giannozzi-jpcm17} with the Perdew-Zunger scheme\cite{perdew-prb81} to treat the exchange correlation energy. The ion cores are described by ultrasoft full-relativistic pseudopotentials, with spin-orbit coupling included, generated using pslibrary 1.0.0,\cite{dalcorso-compmatsci14} with valence configurations 5$d^{10}$6$s^2$6$p^3$ for Bi, 3$s^2$3$p^6$4$s^2$3$d^5$ for Mn, and 2$s^2$2$p^4$ for O. Magnetic order is treated at the non-collinear level. The pseudo-wave functions (pseudo-charge density) are expanded in a plane-wave basis set with kinetic energy cut-off of 90 Ry (450 Ry). Brillouin Zone integrations are performed on a 6$\times$1$\times$4 $\Gamma$-centered Monkhorst-Pack mesh.\cite{monkhorst-prb76} Fractional  electronic occupations near  the Fermi surface are treated with a Gaussian smearing $\sigma$=0.1 eV. The toroidal moment is computed using a custom in-house post-processing code.  The polarization is calculated according to the modified Berry-phase method introduced in a previous study \cite{filippetti-natcomm16} of a ferroelectric metal.

\bibliography{references} 

\begin{thebibliography}{30}
\bibitem [{not()}]{noteM}%
  \BibitemOpen
  \href@noop {} {\bibinfo {title} {\rm $\mathbf{m}_0$ can in principle be
  switched by reversing all the magnetic moments, which would reverse the net
  magnetization as well. \uppercase{H}owever, $\mathbf{m}_0$ identifies the
  additional tilts beyond the ferromagnetic component, thus it can be reversed
  also by changing sign only to the components that are not ferromagnetic.
  \uppercase{I}n this way, $\mathbf{m}_0$ changes sign without switching the
  net magnetization.}}\BibitemShut {Stop}%
\end{thebibliography}
\end{document}